\documentclass[12pt,a4]{article}
\usepackage{amssymb}
\usepackage{amsfonts}
\newcommand{\No}
{{\large\the\textfont2 N}\lower-0.4ex\hbox{\small\underline{$\circ$}}}
\begin{document}
\thispagestyle{empty} 
\begin{center}
{\large\bf M.V. LOMONOSOV MOSCOW STATE UNIVERSITY}\\[3mm] 
{\large\bf D.V. SKOBELTSYN
INSTITUTE OF NUCLEAR PHYSICS}\\ 
\vspace{1.2cm} {\bf Preprint SINP MSU \, 2002--21/705}\\ 
\vspace{2.7cm}
\end{center}

\begin{center}
{\LARGE\bf CONSTRUCTION OF SOLUTIONS FOR\\[5mm]
 THE GENERALIZED H\'ENON--HEILES\\[5mm] SYSTEM
 WITH THE HELP OF\\[5mm] THE PAINLEV\'E TEST}\\[5mm]
 \vspace{1.2cm} 
{\Large\bf S.Yu. Vernov}\\ 
\end{center}

\begin{center}
{e-mail: svernov@theory.sinp.msu.ru}
\end{center}

\vspace{1.2cm}
\begin{abstract}
\normalsize The H\'enon--Heiles system in the general
form has been considered. In a nonintegrable case with the help of the 
Painlev\'e test new solutions have been found as formal Laurent 
or Puiseux series, 
depending on three parameters.  One of parameters determines a location of 
the singularity point, other parameters determine coefficients of  
series. It has been proved, that if absolute values of these two 
parameters are less or equal to unit, then obtained series 
converge in some ring. For some values of these parameters the 
obtained Laurent series coincide with the Laurent series of the known 
exact solutions.
\end{abstract}

\newpage
\section{THE PAINLEV\'E PROPERTY AND INTEGRABI\-LITY} 
A Hamiltonian system in a $2s$--dimensional phase space  
is called { \it completely integrable } (Liouville integrable) if it  
possesses $s$ independent integrals which commute with  
respect to the associated Poisson bracket.  When this is the case, the  
equations of motion are (in principal, at least) separable and solutions  
can be obtained by the method of quadratures. 
 
When we study some mechanical or field theory problem, we imply that  
time and space coordinates are real, whereas the integrability of motion 
equations is connected with the behavior of their solutions as functions of 
complex time and (in the case of the field theory) complex spatial 
coordinates. 
 
S.V.~Kovalevskaya was the first, who proposed~\cite{Kova}  
to consider time as a complex variable and to demand that solutions 
of the motion equations have to be single-valued, meromorphic functions 
on the whole complex (time) plane. 
This idea gave a remarkable result: S.V.~Kovalevskaya discovered a  
new integrable case (nowadays known as the Kovalevskaya's case) for the 
motion of a heavy rigid body about a fixed point~\cite{Kova}  
(see also~\cite{Golubev2,Goriely0}). 
The work of  S.V.~Kovalevskaya   
has shown the importance of application of 
the analytical theory of differential equations to physical problems. 
The essential stage of development of this theory was a classification   
of ordinary differential equations (ODE's) in order of types  
of singularities of their solutions. This classification has  
been made by P.~Painlev\'e.    
 
Let us formulate the Painlev\'e property for ODE's.  
Solutions of a system of ODE's   
are regarded as analytic functions, may be  
with isolated singular points~\cite{Golubev1, Hille}.  
A singular point of a solution is said {\it critical } (as opposed to    
{\it noncritical}) if the solution is multivalued (single-valued) 
in its neighborhood and {\it  movable} if its location  
depends on initial conditions\footnote{ Solutions of a system with a 
time-independent Hamiltonian can have only movable singularities.}.

\vspace{7.2mm} 
 
{\large\it Definition. } {\it A system of  ODE's has  
 \textbf{\textit{ the Painlev\'e 
property }} if its general solution has no movable  
critical singular point}~\cite{Painleve1,Painleve2}. 
 
\vspace{7.2mm}

An arbitrary solution of such system is single-valued in the neighborhood  
of its singular point $t_0$ and can be expressed as a Laurent series with 
a finite number of terms with negative powers of $t-t_0$.  
If a system has not the Painlev\'e 
property, but, after some change of variables, the obtained system  
possesses this property, then the initial system is said to  
have \textit{the weak Painlev\'e property}.  
 
Investigations of many dynamical systems,  
Hamiltonian~[8--10] or dissipative (for example,  
the Lorenz systems~[10--13]), show, that  
a system is completely integrable only for such values of parameters,  
at which it has the Painlev\'e property (or the weak  Painlev\'e  
property).  Arguments, which clarify the connection between the Painlev\'e  
analysis and the existence of motion integrals, are presented  
in~\cite{ES1,ES2}.  If the system misses the Painlev\'e property (has  
complex or irrational "resonances"), then the system cannot be  
"algebraically integrable"~\cite{Y1} (see also~\cite{Y2} and references  
therein).  At the same time the integrability of an arbitrary system  
with the Painlev\'e property has yet to be proved.  There is not an  
algorithm for construction of the additional integral by the Painlev\'e  
analysis. It is easy to give an example of an integrable system without  
the Painlev\'e property~\cite{Kozlov}:  $H=\frac{1}{2}p^2+f(x)$, where  
$f(x)$ is a polynomial which power is not lower than five. The given  
system is trivially integrable, but its general solution is not a  
meromorphic function.  
 
The Painlev\'e test is any algorithm designed to determine necessary 
conditions for a differential equation to have the Painlev\'e property. 
The original algorithm, developed by P.~Painlev\'e and 
used by him to find all the second order ODE's with  
Painlev\'e property~\cite{Painleve2}, is known as the $\alpha$-method.   
The method of S.V.~Kovalevskaya is not as general as the  
$\alpha$--method, but much more simple\footnote{ Different variants 
of the Painlev\'e test are compared in~\cite[R.~Conte  
paper]{Conte0}}.  
 
In 1980, motivated by the work of S.V.~Kovalevskaya~\cite{Kova}, 
 \  M.J.~Ablowitz, A.~Ramani and H.~Segur~\cite{ARS}  
developed a new algorithm of the Painlev\'e test for ODE's.  
The remarkable property of this test is that it can be checked in a 
finite number of steps. They also were the first to point out the  
connection between the nonlinear partial differential equations (PDE's),  
which are soluble by the inverse scattering transform method, and ODE's  
with the Painlev\'e property. Subsequently the Painlev\'e property for PDE  
was defined and the corresponding Painlev\'e test (the WTC procedure) was  
constructed~\cite{WTC,Weiss4} (see  also~\cite[23--29]{Conte0}). With the  
help of this test it has been found, that all PDE's, which are solvable by  
the inverse scattering transforms, have the Painlev\'e property, may be,  
after some change of variables.  For many integrable PDE's, for example,  
the Korteweg--de-Vries~\cite{Tabor} and the 
sine--Gordon~\cite{Weissine} equations, the 
B\"acklund transformations and the Lax representations result from the WTC 
procedure~\cite{Weiss4, Weiss5, ConteTMF}.  Also, special 
solutions for certain nonintegrable PDE's were constructed using this 
algorithm~\cite{Tabor2,Conte3}. 
 
The algorithm for finding  
special solutions for ODE system in the form of a finite expansion in  
powers of unknown function $\varphi(t-t_0)$ has been  
constructed in~\cite{Weiss1,Weiss2}.  The function $\varphi(t-t_0)$ and  
coefficients have to satisfy some system of ODE, often more simple than  
an initial one. This method has been used~\cite{Sahadevan} to construct  
 exact solutions for certain nonintegrable systems of ODE's.  
 
 The aim of this paper is to find new special solutions  
for the generalized H\'enon--Heiles system using the Painlev\'e test.   
In distinction to~\cite{Sahadevan} we obtain solutions as formal  
Laurent or Puiseux series and find domains of their convergence.

\section{THE H\'ENON--HEILES HAMILTONIAN} 
 
Let us consider the motion of a star in an axial-symmetric and  
time-inde\-pendent potential.  The motion equations admit two  
well-known integrals (energy and angular momentum) and would  
 be solved by the method of quadratures if the third integral of motion 
 is known. Due to the symmetry of the potential the considered system 
is equivalent to two-dimensional one. However, for many  
polynomial potentials the obtained system has not 
the second integral as a polynomial function.  
 
In the 1960s, asymptotic methods~\cite{Contop,Gustavson}  
have been developed to show either existence or absence of the third  
integral for some polynomial potentials.  
To answer the question about the existence of the third integral 
H\'enon and Heiles~\cite{HeHe} considered the behavior of numerically 
integrated trajectories. Emphasizing that their   
choice of potential does not proceed from experimental data,  
H\'enon and Heiles have proposed the following Hamiltonian:   
 
$$  
H=\frac{1}{2}\Big(x_t^2+y_t^2+ x^2+y^2\Big)+x^2y-\frac{1}{3}y^3,  
\eqno(1) 
$$  
because: on the one hand, it is 
analytically simple; this makes the numerical computations of trajectories 
easy;  on the other hand, it is sufficiently complicated to 
 give trajectories which are far from trivial. Indeed, for low energies  
the  H\'enon--Heiles system appears to be  
integrable, in so much as trajectories (numerically integrated) always lay  
on well-defined two-dimensional surfaces. On the other hand,  
for high energies many of these integral surfaces are  
destroyed, it points on absence of the third integral.  
 
Subsequent numerical investigations~\cite{ChTW1,ChTW2}  show,  
that in the complex 
 $t$-plane singular points of solutions of the motion  
equations group in self-similar spirals. It turns out extremely  
complicated distributions of singularities, forming a boundary,  
across which the solutions can not be analytically continued.  
 
The generalized H\'enon--Heiles system is described by the Hamiltonian:   
 
$$  
H=\frac{1}{2}\Big(x_t^2+y_t^2+\lambda  
x^2+y^2\Big)+x^2y-\frac{C}{3}\:y^3\eqno(1') 
$$ 
 
and the corresponding system of the motion equations:  
 
$$  
\left\{  
\begin{array}{lcl} x_{tt} & = &-\lambda x -2xy,\\[2mm] y_{tt} & = &-y  
-x^2+Cy^2, \end{array} \right.  \eqno(2)  
$$  
where 
$x_{tt}\equiv\frac{d^2x}{dt^2}$ and $y_{tt}\equiv\frac{d^2y}{dt^2}$, \   
$\lambda$ and $C$ are numerical parameters.  
 
Due to the Painlev\'e analysis the following integrable cases of~$(2)$ 
have been found:   
 
$$ \begin{array} {cll} \mbox{(i)} & C=-1,  
&\lambda=1,\\ \mbox{(ii)} & C=-6, &\mbox{$\lambda$ is an arbitrary  
number},\\ \mbox{(iii)} & C=-16,\quad &\lambda=\frac{1}{16}.\\ \end{array}  
$$ 
 
\vspace{2.7mm} 
 
In contradiction to the case (i) the cases (ii) and (iii) 
are nontrivial, so the integrability of these cases had to be proved 
additionally. In the 1980's the required second integrals were  
constructed~[40--43].  For integrable cases of the  H\'enon--Heiles 
system the B\"acklund transformations~\cite{Weiss1,Weiss2} and the Lax 
representations~\cite{NTZ,FNT,Polska} have been found.  

The three integrable cases of the 
H\'enon--Heiles system correspond precisely to the stationary flows of the 
only three integrable cases of {\it fifth-order polynomial nonlinear 
evolution} equations of scale weight 7 (respectively the Sawada--Kotega, 
the fifth-order Korteweg--de Vries and the Kaup--Kupershmidt 
equations)~\cite{Weiss2,Fordy2,Polska}.
 
The  H\'enon--Heiles system is a model widely used in physics,  
in particular, in gravitation~[46--48] and plasma theory~\cite{Guo}.   
 The models, described by the Hamiltonian $(1')$ with some additional  
nonpolynomial terms, are actively studied~\cite[50--52]{Polska} as well.

\section{NONINTEGRABLE CASES} 
 
The general solutions of the H\'enon--Heiles system   
are known only in integrable cases~\cite{Conte4}, in other cases search of  
new (exact or asymptotic) solutions is an actual problem.  
 
The procedure for transformation the Hamiltonian to a normal form  
and for construction the second independent integral in the form of 
formal power series in the phase variables $x$, $x_t$, $y$ and $y_t$  
(Gustavson integral) has been realized for  
the H\'enon--Heiles system both in the original ($\lambda=1$, 
$C=1$)~\cite{Gustavson} (see also~\cite{Moser}) and in the general 
forms~\cite{Braun,Kasper}. Using the Bruno algorithm~\cite{Bruno1,Bruno2}  
\ V.F.~Edneral \  has constructed the Poincar\'e--Dulac normal form and  
found~\cite{Edneral1,Edneral2} (provided that all phase variables are  
small) local families of periodic solutions.  Recently it has been  
found that a local series around the singularities in the complex (time)  
plane can be transformed to some local series around the singularities at  
the fixed points in phase space and analyzed via normal forms  
theory~\cite{Goriely1,Goriely2}.

The H\'enon--Heiles system as a system of two second order ODE's 
is equivalent to the fourth order equation\footnote{ For given $y(t)$ the  
function $x^2(t)$ is a solution of a linear equation. System $(2)$ is  
invariant to exchange $x$ to $-x$. }:   
$$  
y_{tttt}=(2C-8)y_{tt}y -  
(4\lambda+1)y_{tt}+2(C+1)y_{t}^2+frac{20C}{3}y^3+  
(4C\lambda-6)y^2-\lambda y-4H,  \eqno(3)  
$$ 
where $H$ is the energy of the system. 
 
To find a special solution of 
the given equation one can assume that $y$ satisfies some more simple 
equation.  For example, the well-known solutions 
in terms of the Weierstrass elliptic functions~\cite{BE,Gera} satisfy the 
following first-order differential equation:  
$$ 
  y_t^2=\tilde {\cal A}y^3+\tilde {\cal B}y^2+\tilde {\cal C} y+ 
\tilde  {\cal D}, \eqno(4) 
$$ 
where  
$$ 
\tilde {\cal A}=\frac{2}{3}C,\qquad 
\tilde {\cal B}=-1,\qquad \tilde {\cal C}=0\quad\mbox{and}\quad  
\tilde {\cal D}=2H \eqno(4a) 
$$ 
or 
$$ 
\begin{array}{lcl} 
\displaystyle \tilde {\cal A}&\displaystyle =&\displaystyle  
-\:\frac{4}{3},\\[7mm]\tilde {\cal B}&\displaystyle  
=&\displaystyle \frac{1-(C+2)\lambda}{C+1},\\[7mm] \displaystyle \tilde  
{\cal C}&\displaystyle =& \displaystyle -\:\frac{3C^2\lambda^2 -  
3C^2\lambda + 8C\lambda^2 - 7C\lambda - C + 4\lambda^2 - 2\lambda -  
2}{3C^3 + 10C^2 + 11C + 4}, \\[7mm] \displaystyle \tilde {\cal  
D}&\displaystyle =&\displaystyle \frac{24C^4H + 104C^3H - 9C^3\lambda^3 +  
6C^3\lambda^2 + 3C ^3\lambda } 
{4(3C^5 + 22C^4 + 60C^3 + 78C^2 + 49C + 12)}\: +\\[5mm]  
&\displaystyle  + &\displaystyle \frac{168C^2H -  
30C^2\lambda^3+13C^2\lambda^2 + 16C^2 \lambda+C^2}{4(3C^5 +  
22C^4 + 60C^3 + 78C^2 + 49C + 12)}\:+ \\[5mm] 
&\displaystyle  + & \displaystyle \frac{120CH - 28C\lambda^3 +  
24C\lambda+4C + 32H - 8\lambda^3 - 4\lambda^2 + 8\lambda + 4}{4(3C^5 +  
22C^4 + 60C^3 + 78C^2 + 49C + 12)}.  
\end{array}\eqno(4b)  
$$  
$\tilde {\cal  
 D}$ is proportional to energy $H$ (arbitrary parameter),  
therefore, solutions  
$(4a)$ and $(4b)$ are two-parameter ones. 
 
E.I. Timoshkova~\cite{Timosh} generalized equation $(4)$: 
$$ 
  y_t^2=\tilde {\cal  A}y^3+\tilde {\cal B} y^2+\tilde {\cal C} y+ 
\tilde {\cal D}+\tilde {\cal G} y^{5/2}+\tilde  {\cal E} y^{3/2} 
\eqno(5) 
$$ 
and found new one-parameter sets of solutions of the  
H\'enon--Heiles  system in nonintegrable cases 
($C= -\:\frac{4}{3}$ \ or \ $C= -\:\frac{16}{5}$, \   
$\lambda$ is an arbitrary number).  
These solutions (i.e. solutions with $\tilde {\cal G}\neq 0$ or  
$\tilde {\cal E} \neq 0 $) are derived only at $\tilde {\cal D}=0$,  
therefore, substitution $y=\varrho^2$ gives:   
$$  
\varrho_t^2=\frac{1}{4}\Bigl(\tilde {\cal  
A}\varrho^4+\tilde {\cal G} \varrho^3+\tilde {\cal B} \varrho^2 +   
\tilde {\cal E}\varrho+\tilde {\cal C}\Bigr).  \eqno (6)  
$$ 
 
The general solution of~$(6)$ has one arbitrary parameter  
and can be expressed in elliptic functions.   
 
In this paper I analyze system~$(2)$ at $C = -\:\frac{16}{5}$ and   
$\lambda=\frac{1}{9}$ ({\it the Solution 2.2} of the  
paper~\cite{Timosh}). In this case equation $(5)$ is:   
$$  
y_t^2+\:\frac{32}{15}y^3+\frac{4}{9}y^2\pm\frac{8i}{\sqrt{135}}y^{5/2}=0 
\eqno (7) 
$$ 
and, depending on a choice of a sign before the last term, we obtain  
 either (in case of  sign $ + $): 
$$  
  y=-\:\frac{5}{3\left(1-3\sin\left(\frac{t-t_0}{3}  
\right)\right)^{2^{\vphantom{27}}}} \qquad\mbox{and}\qquad  
  x^2=\frac{25\Big(1-\sin\left(\frac{t-t_0}{3}\right)\Big)} 
{9\left(1-3\sin\left(\frac{t-t_0}{3}\right)\right)^{3^{\vphantom{27}}}}; 
\eqno (8.1) 
$$ 
or (in case of  sign $ - $): 
$$  
  y=-\:\frac{5}{3\left(1+3\sin\left(\frac{t-t_0}{3}  
\right)\right)^{2^{\vphantom{27}}}}  \qquad\mbox{and}\qquad  
  x^2=\frac{25\Big(1+\sin\left(\frac{t-t_0}{3}\right)\Big)} 
{9\left(1+3\sin\left(\frac{t-t_0}{3}\right)\right)^{3^{\vphantom{27}}}}. 
\eqno (8.2) 
$$

\section{RESULTS OF THE  PAINLEV\'E TEST FOR THE  
H\'ENON--HEILES  SYSTEM}  
 
The Ablowitz--Ramani--Segur algorithm of the  Painlev\'e test 
appears very useful to find asymptotic solutions as a formal Laurent series. 
 
We assume that the behavior of solutions in a sufficiently small 
neighborhood of the singularity is algebraic, it means that  
$x$ and $y$ tend to infinity as some powers of $t-t_0$: 
$$  
x=a_{\alpha}(t-t_0)^\alpha\qquad \mbox{and}\qquad  
y=b_{\beta}(t-t_0)^\beta,\eqno(9)  
$$  
where $\alpha$, $\beta$, $a_{\alpha}$ and $b_{\beta}$ are some constants. 
We assume that real parts of $\alpha$ and $\beta$ are less then zero, 
and, of course, $a_{\alpha}\neq 0$ and $b_{\beta}\neq 0$.

If $\alpha$ and $\beta$ are integer numbers, then  substituting  
$$  
x=a_{\alpha}(t-t_0)^{\alpha}+\sum\limits_{j=1}^{N_{max}}  
a_{j+\alpha}(t-t_0)^{j+\alpha},\eqno (10.1) 
$$ 
$$ 
y=b_{\beta}^{\vphantom{27}}(t-t_0)^{\beta}+ 
\sum\limits_{j=1}^{N_{max}} 
b_{j+\beta}(t-t_0)^{j+\beta}\eqno (10.2) 
$$ 
one can transform the ODE system into a set of linear  
algebraic systems in coefficients $a_{k}$ and  $b_{k}$.  
In the general case one can obtain the exact  
solutions (in the form of formal Laurent series) only if one solves  
infinity number of systems ($N_{max}=\infty$). On the other hand,  
if one solves a finite number of systems one obtains asymptotic solutions.   
With the help of some computer algebra system, for example, the system {\bf  
REDUCE}~\cite{REDUCE,4444},  
these systems can be solved step by step and  
asymptotic solutions can be automatically found with any accuracy. But  
previously one has to determine values of constants $\alpha$, $\beta$,  
$a_\alpha^{\vphantom{27}}$ and  
$b_\beta^{\vphantom{27}}$ and to analyze 
systems with zero determinants. Such systems 
correspond to new arbitrary constants or have no solutions. 
Powers at which new arbitrary constants enter are called 
{\it resonances}. The Painlev\'e test gives all information about  
possible dominant behaviors and resonances (see, for example,
\cite{Tabor}). Moreover, the results of the  Painlev\'e analysis 
point out cases, in which it is useful to include into expansion terms 
with fractional powers of  $t-t_0$.    
 
For the generalized H\'enon-Heiles system  
there exist two possible dominant behaviors and resonance  
structures~\cite{Tabor,ChTW2}: 
 
\vspace{2.72mm} 
\begin{tabular}{|l|l|} 
\hline 
{ \it Case 1}:  &  { \it Case 2}: ($\beta<\Re e(\alpha)<0$)\\[2mm] 
\hline 
 $\alpha=-2$,& $\alpha=\frac{1\pm\sqrt{1-48/C}^{\vphantom{7^4}}}{2}$,\\[1mm] 
$\beta=-2$, & $\beta=-2$,\\[1mm] 
$a_{\alpha}=\pm 3\sqrt{2+C}$, &  
$a_{\alpha}=c_1^{\vphantom{4}}$ {\bf (an arbitrary number)},\\[1mm] 
$b_{\beta}^{\vphantom{27}}=-3$,  
& $b_{\beta}^{\vphantom{27}}=\frac{6}{C}$, \\[1mm] 
$r=-1,\; 6,\; \frac{5}{2}\pm \frac{\sqrt{1-24(1+C)}}{2}$. &  
$r=-1,\; 0,\; 6,\; \mp\sqrt{1-48/C}$.\\[1mm] 
\hline 
\end{tabular} 
\vspace{2.72mm} 
 
In the Table the values of $r$ denote resonances: $r=-1$ corresponds to 
arbitrary parameter $t_0$; $r=0$ (in the {\it Case 2}) corresponds to 
arbitrary parameter $c_1^{\vphantom{4}}$.  
Other values of $r$ determine powers of $t$, to be exact,  
$t^{\alpha+r}$ for $x$ and $t^{\beta+r}$ for $y$, at which new  
arbitrary parameters enter (as solutions of systems with zero determinants).  
 
For integrability of system $(2)$ all values of $\alpha$  
and  $r$ have to be integer (or rational) and all  
systems with zero determinants have to have solutions at all  
values of included in them free parameters. It is possible only in  
the cases (i) --- (iii).  
  
At $C=-2$ (in the {\it  Case 1}) \ $ a_{\alpha}=0$. It is 
the consequence of the fact that, contrary to our assumption, the  
behaviour of the solution in the neighborhood of a singular point is not  
algebraic, because its dominant term includes logarithm.

Those values of $C$, at which $\alpha$ and $r$  
are integer (or rational) numbers either only in the {\it  Case 1} or only  
in the {\it  Case 2}, are of interest for search of special solutions.

\section{NEW SOLUTIONS}  
 
\subsection{Finding of solutions in the form of formal Laurent series}  
 
Let us consider the  H\'enon--Heiles system  
with $C=-\:\frac{16}{5}$.  In the {\it Case 1} some  values  of $r$ are  
not rational, so it is a nonintegrable system.  To find special  
asymptotic solutions let us consider the {\it Case 2}.  In this case  
$\alpha=-\:\frac{3}{2}$ and $r=-1,\;0,\;4,\;6$, hence, in the neighborhood  
of the singular point $t_0$ we have to seek $x$ in such form that $x^2$  
can be expand into Laurent series, beginning from $(t-t_0)^{-3}$.   
Let $t_0=0$, substituting  
$$  
x=\sqrt{t}\left(c_1^{\vphantom{4}}t^{-2}+\sum_{j=-1}^\infty a_jt^j\right)  
\qquad\mbox{and}\qquad y=-\:\frac{15}{8}t^{-2}+\sum_{j=-1}^\infty b_jt^j  
$$  
in $(2)$, we obtain the following sequence of linear system in 
 $a_k$ and $b_k$:   
$$  
\left\{ 
\begin{array}{l} 
\displaystyle\Big  
(k ^ 2-4\Big)a_{k} + 2c_1^{\vphantom{4}} b_{k} = -\lambda a_{k-2}  
-2\sum_{j=-1}^{k-1}a_{j}b_{k-j-2}\\[7.2mm]  
\displaystyle\Big((k-1)k-12\Big)b_{k}  
= -\:b_{k-2} - \sum_{j=-2}^{k-1}a_{j}a_{k-j-3} -  
\frac{16}{5}\sum_{j=-1}^{k-1} b_{j}b_{k-j-2}.  
\end{array} 
\right. 
\eqno (11)  
$$  
 
 If $k=2$ or $k=4$, then the determinant of $(11)$ is equal 
to zero. To determine $a_2$ and $b_2$ we have the following system:   
$$  
\left\{  
\begin{array}{l@{}}  
\displaystyle c_1^{\vphantom{4}}\Big(557056 c_1^8 + \big(15552000\lambda -  
4860000\big)c_1^4 + 864000000b_2 \: + \\[1mm]\displaystyle   +\:  
108000000\lambda^2 - 67500000\lambda +10546875\Big)=0,\\[2mm]\displaystyle   
 818176c_1^8  + \Big(15660000\lambda - 4893750\Big)c_1^4 -  
\\[1mm]\displaystyle  -\: 810000000b_2- 6328125=0.   
\end{array}  
\right.   
\eqno (12)  
$$ 
 
As one can see this system does not include terms, which are proportional 
to $a_2$, hence, $a_2$ is an arbitrary parameter (a constant of integration). 
 
We discard the solution with $c_1^{\vphantom{4}}=0$ and obtain   
the system in $\tilde c_1^{\vphantom{4}}\equiv c_1^4$ and  $b_2$  
with the following solutions:   
$$  
\begin{array}{rcl}  
\displaystyle \tilde  
c_1^{\vphantom{4}}&\displaystyle =&\displaystyle  
\frac{1125(4\sqrt{35(2048\lambda^2 - 1280\lambda +  
387)} - 1680\lambda + 525)}{16755 2}, \\[2.7mm]  
\displaystyle b_2& \displaystyle  =&\displaystyle  - \frac{(10944\lambda - 3420)\sqrt{35(2048\lambda^2 -  
1280\lambda + 387)}- 4403456\lambda^2 + 2752160\lambda - 789065}{117956608}  
\end{array} 
$$  
or  
$$  
\begin{array}{rcl} 
\displaystyle \tilde c_1^{\vphantom{4}}&\displaystyle  = 
 &\displaystyle \frac{1125( -  
4\sqrt{35(2048\lambda^2 - 1280\lambda + 387)} -  
1680\lambda + 525)}{167552}, \\[2.7mm] 
\displaystyle b_2 &\displaystyle  =&\displaystyle  \frac{(10944\lambda  
- 3420)\sqrt{35(2048\lambda^2 - 1280\lambda + 387)} -  
4403456\lambda^2 + 2752160\lambda - 789065} {117956608}.   
\end{array} 
$$  
 
We obtain new constant of integration $a_2$,  
but we must fix $c_1^{\vphantom{4}}$, so number of  
constants of integration is equal to 2.  It is easy to verify that  
$b_4$ is an arbitrary parameter, because the  corresponding  
system is equivalent to one linear equation.  So,  
using Painlev\'e test, we obtain an asymptotic  
solution which depends on three parameters, namely  $t_0$,  $a_2$ and $b_4$. 
 
Now asymptotic solutions can be obtained with arbitrary accuracy.  
For given $\lambda$ one has to choose  
$c_1^{\vphantom{4}}$ as one of the roots of system~$(12)$. After this the  
coefficients $a_j$ and $b_j$ can be found automatically  
with the help of some computer algebra system.   
 
 For example, if $\lambda=\frac{1}{9}$, then $(12)$ has the following  
solutions:   
$$  
\left\{\tilde c_1^{\vphantom{4}}=\frac{625}{128},   
\quad b_2=-\:\frac{1819}{663552}\right\}, \qquad\left\{\tilde  
  c_1^{\vphantom{4}}=-\:\frac{8125}{23936},  \quad  
b_2=-\:\frac{8700683}{1364926464}\right\}. 
$$ 
 
Taking into account, that system $(2)$ is invariant to change $x$ to  
$-x$, we obtain four types of formal solutions:  
$$  
\begin{array}{@{}l@{}}  
\displaystyle x =  
\sqrt{t}\left\{\frac{5\sqrt[4]{2}}{4}t^{-2}  
+\frac{25}{96\sqrt[4]{2}}t^{-1} 
-\frac{5\sqrt[4]{2}}{16}+\frac{5275}{663552\sqrt[4]{2}}t\: 
+a_2t^2\dots\right \}, \\[7.2mm] 
\displaystyle y=  
-\:\frac{15}{8}t^{-2}+\frac{5\sqrt{2}}{32}t^{-1}-\frac{205}{2304}+ 
\frac{115\sqrt{2}}{13824}t-\frac{1819}{663552}t^2\:+  
\\[5mm] 
 \displaystyle + 
\left(\frac{741719\sqrt{2}}{1528823808}  
+\frac{5\sqrt[4]{2}}{12}a_2\right)t^3+b_4t^4+\dots; 
\end{array}\eqno(13.1) 
$$ 
$$ 
\begin{array}{@{}l@{}} 
\displaystyle x= \sqrt{t}\left\{\frac{5i\sqrt[4]{2}}{4}t^{-2} 
-\frac{25i}{96\sqrt[4]{2}}t^{-1} 
-\frac{5i\sqrt[4]{2}}{9216} 
- \frac{5275i}{663552\sqrt[4]{2}}t^1+a_2t^2+\dots\right\}, \\[7.2mm]  
\displaystyle y=  
-\:\frac{15}{8}t^{-2}-\frac{5\sqrt{2}}{32}t^{-1}-\frac{205}{2304}- 
\frac{115\sqrt{2}}{13824}t-\frac{1819}{663552}t^2\:-  
\\[5mm] 
\displaystyle -  
\left(\frac{741719\sqrt{2}}{1528823808}+ 
\frac{5i\sqrt[4]{2}}{12}a_2\right)t^3+ 
b_4t^4+\dots; 
\end{array}\eqno(13.2) 
$$ 
$$ 
\begin{array}{@{}l@{}} 
\displaystyle x=\sqrt{t}\left\{\frac{5\sqrt{2}} 
{4}\sqrt[4]{-\:\frac{13}{374}}t^{-2} 
+\frac{25i\sqrt{2431}}{17952}\sqrt[4]{-\:\frac{13}{374}}t^{-1}\: 
- \right.\\[4mm] 
\displaystyle  -  
\left.\frac{38645\sqrt{2}}{574464}\sqrt[4]{-\:\frac{13}{374}}- 
\frac{7028575i\sqrt{2431}}{23203749888} 
\sqrt[4]{-\:\frac{13}{374}}t+a_2t^2+\dots\right\}, \\[7.2mm]  
\displaystyle y=-\:\frac{15}{8}t^{-2}+ 
\frac{5i\sqrt{4862}}{5984}t^{-1}-\frac{69335}{430848}- 
\frac{37745i\sqrt{4862}}{483411456}t-\frac{8700683}{1364926464}t^2\:-  
\\[4mm] \displaystyle -    
\left(\frac{1148020763i\sqrt{13}\sqrt{374}}{3332429743915008}  
-\frac{5\sqrt{2}}{12}a_2\sqrt[4]{-\:\frac{13}{374}}\right)t^3+b_4t^4+  
\dots; 
\end{array} \eqno(13.3)
$$ 
 
$$ 
\begin{array}{@{}l@{}} 
\displaystyle x=  
 \sqrt{t}\left\{\frac{5i\sqrt{2}}{4}\sqrt[4]{-\:\frac{13}{374}}t^{-2} 
+\frac{25\sqrt{2431}}{17952}\sqrt[4]{-\:\frac{13}{374}}t^{-1}\: 
-\right.\\[4mm] 
\displaystyle  -  
\left.\frac{38645i\sqrt{2}}{574464}\sqrt[4]{-\:\frac{13}{374}}- 
\frac{7028575\sqrt{2431}}{23203749888} 
\sqrt[4]{-\:\frac{13}{374}}t+a_2t^2+\dots\right\},\\[5mm]  
\displaystyle y= 
-\:\frac{15}{8}t^{-2}-\frac{5i\sqrt{4862}}{5984}t^{-1}-\frac{69335}{430848}+ 
\frac{37745i\sqrt{4862}}{483411456}t-\frac{8700683}{1364926464}t^2\:-  
\\[4mm]  
 \displaystyle -  
\left(\frac{1148020763\sqrt{13}\sqrt{374}}{3332429743915008}  
+\frac{5i\sqrt{2}}{12}a_2\sqrt[4]{-\:\frac{13}{374}}\right)t^3+b_4t^4+ \dots. 
\end{array} \eqno(13.4)
$$

It is easy to verify that if 
$$  
 a_2=-\:\frac{21497\sqrt[4]{2}} 
 {42467328}\qquad\mbox{and}\qquad  
 b_4=-\:\frac{858455}{12039487488},  
$$  
then series $(13.1)$ are the Laurent series of $(8.1)$. Also if  
$$  
 a_2=-\:\frac{21497i\sqrt[4]{2}} 
 {42467328}\qquad\mbox{and}\qquad  
 b_4=-\:\frac{858455}{12039487488},  
$$ 
then series $(13.2)$ are the Laurent series of $(8.2)$.

\subsection{Convergence of the obtained series}  
When an asymptotic series is obtained  
the question about its convergence arises. It is known that 
a domain of Laurent series convergence is a ring.  
Let us find conditions, at which the obtained 
series converge in the following ring:  
$ 0<|t|\leqslant 1-\varepsilon $, where $ \varepsilon $ is  
any positive number.  
 
The sum of a geometrical progression  
$\displaystyle  
S = \sum\limits_{n = 0}^\infty t^n = \frac {1}{1-t}$  
is finite if  $|t|\leqslant 1-\varepsilon$, hence, our series  
converge in the above-mentioned ring, if $\exists N$ such  
that $\forall n> N$ \ $|a_n| \leqslant 1$ and $|b_n| \leqslant 1$. 
 
Let $|a_n|\leqslant 1$ and $|b_n|\leqslant 1$ for all $-1 \leqslant n < k$, 
then from $(11)$, we obtain:   
$$  
|a_k|\leqslant\frac{2k+2+|\lambda| + 2|c_1^{\vphantom{4}}|}{|k^2-4|},  
\qquad |b_k|\leqslant\frac{21(k + 2)}{5(k^2-k-12)}. \eqno (14)  
$$ 
 
It is easy to see that there exists such $N$, that, if 
$|a_n|\leqslant 1$ and $|b_n|\leqslant 1$ for $-1\leqslant n\leqslant N$,  
then $|a_n | \leqslant 1$ and $ |b_n| \leqslant 1$ for $-1\leqslant n <  
\infty$. $N$ is a  maximum from $8$ and  
$1+\sqrt{|\lambda| + 2 |c_1^{\vphantom{4}}|+7}$. 
                                                  
For example, if $\lambda=\frac{1}{9}$, then for any possible value  
of $c_1^{\vphantom{4}}$, we obtain $N=8$. It is easy to verify that 
if $|a_2|\leqslant 1$ and $|b_4|\leqslant 1$, then  
$ |a_n|\leqslant 1$ and $|b_n| \leqslant 1$ for  
$-1\leqslant n\leqslant  
8$, and, hence, for an arbitrary $n$.  Thus our Laurent series converge   
in the ring $ 0 < | t | \leqslant 1-\varepsilon$.  
Numerical analysis shows~\cite{Vernov} that these series can also converge  
at absolute values of parameters more than unit.  
For other values of $\lambda$ the convergence can be considered analogously. 
 
\section{GENERALIZATION OF SOLUTIONS IN TERMS OF THE WEIERSTRASS ELLIPTIC 
FUNCTIONS} 
 
Let us consider solutions $(4)$. The values of parameters $(4a)$ 
correspond to $x(t)\equiv 0$. Solutions $(4b)$ correspond to 
the {\it Case 1} (see Table) and $x(t)$ can be expressed 
in terms of the Weierstrass elliptic functions. 
For some values of $C$ these two-parameter solutions  can be  
generalized.  
For example, if $C=-\:\frac{9}{8}$, then some resonances are 
half-integer. Substituting 
$$ 
  x=\sum_{k=-4}^{\infty}\tilde a_kt^{k/2}\qquad\mbox{and}\qquad  
 y=\sum_{k=-4}^{\infty}\tilde b_kt^{k/2}   
$$ 
in $(2)$, we obtain that $\tilde a_n$ and $\tilde b_n$ have to satisfy 
the following system:  
$$  
\left\{ 
\begin{array}{l} 
\displaystyle \frac{n(n-2)-24}{4}\tilde a_n 
+2\tilde  a_{-4}\tilde b_n=  
-\lambda \tilde a_{n-4}-2\sum\limits_{k=-1}^{n-1}\tilde a_{k} 
\tilde b_{n-k-4},\\[7.2mm]  
\displaystyle 2\tilde a_{-4}\tilde a_n+\frac{n(n-2)-27}{4}\tilde b_n =  
 -\:\tilde b_{n-4} - \sum\limits_{k=-3}^{n-1}\tilde a_{k}\tilde a_{n-k-4} -  
\frac{9}{8}\sum\limits_{k=-3}^{n-1}\tilde  b_{k}\tilde b_{n-k-4},   
\end{array}  
\right.   
$$ 
 where $\tilde a_{-4}=\pm\:\frac{3\sqrt{7}}{2\sqrt{2}}$ and $\tilde  
b_{-4}=-3$. 
 
At any $\lambda$ we obtain three-parameter solutions as  formal  
Puiseux series. For example, if $\lambda=1$  
(and $\tilde a_{-4}=\frac{3\sqrt{7}}{2\sqrt{2}}$)  
the solution is the following ($t_0=0$): 
$$ 
\begin{array}{@{}rcl@{}} 
x\:&=&\frac{3\sqrt{7}}{2\sqrt{2}}t^{-2} 
+\frac{7}{8\sqrt{2}}+\frac{4}{\sqrt{14}}D_1t^{3/2} 
+\frac{\sqrt{7}}{160\sqrt{2}}t^2-   
\frac{15\sqrt{7}}{224\sqrt{2}}D_1t^{7/2}\:-\\[2.7mm] 
&-& 
\frac{\sqrt{7}}{2\sqrt{2}}D_2t^4 
- \frac{467\sqrt{7}}{8624\sqrt{2}}D_1^2t^5  
+ \frac{1157\sqrt{7}}{430080\sqrt{2}}D_1t^{11/2} 
+ \frac{\sqrt{7}}{115200\sqrt{2}}t^6+\dots,\\[5mm]  
y\:&=&\: 
-3t^{-2}-\frac{1}{4}+D_1t^{3/2}-\frac{1}{80}t^2-  
\frac{15}{128}D_1t^{7/2}\:+ \\[2.7mm] 
 &+& 
 D_2t^4- \frac{79}{616}D_1^2t^5+\frac{1157}{245760}D_1t^{11/2} 
\frac{1}{57600}t^6+ \dots.   
\end{array} \eqno(15) 
$$  
 
Using numerical calculations it is easy to show that if  
$|D_1|<1$ and  $|D_2|<1$ then 
 $|\tilde a_{k}|<1$ and $|\tilde b_{k}|<1$,  
for $-3\leqslant  k\leqslant  50$, except only  
$\tilde a_{3}=\sqrt{\frac{8}{7}}D_1$. It is  
sufficient to prove that  $|\tilde a_{k}|<1$ and $|\tilde b_{k}|<1$ for  
all $ k > 50$ and, hence, our series converge in the ring $ 0<|t|\leqslant  
1-\varepsilon$. 
If $D_1=0$ then $y$ satisfies $(4)$ with   
$$ 
\tilde {\cal A}=-\:\frac{4}{3},\qquad 
\tilde {\cal B}=-1,\qquad \tilde {\cal C}=0\quad\mbox{and}\quad  
\tilde {\cal D}=\frac{16}{15}\:H, \eqno(4b') 
$$ 
and solution can be presented in terms of  
the Weierstrass elliptic functions.

\section{CONCLUSION} 
 
Using the Painlev\'e analysis one can not only find   
integrable cases of dynamical systems, but also construct  
special solutions in nonintegrable cases. 
 
We have found the special solutions of the H\'enon--Heiles system 
with $C=-\:\frac{16}{5}$ as formal Laurent series, depending on three  
parameters. For some values of two parameters the obtained solutions  
coincide with the known exact solutions.  
At  $C=-\:\frac{9}{8}$ two-parameter solutions in terms of the  
Weierstrass elliptic functions have been generalized to three-parameter  
ones. New solutions found as formal Puiseux series. 
For some values of $\lambda$ the analysis of convergence of  
the obtained series has been made and it has been proved, that they have  
nonzero domain of convergence. Similar analysis can been made  
for any value of $\lambda$.  
  
With the help of the Painlev\'e test particular  
asymptotic solutions as Laurent or Puiseux series can be found for 
the H\'enon--Heiles system with some other values of $C$ and $\lambda$.  
Just at these values of parameters the probability of   
finding of new exact solutions similar to the solutions found  
in~\cite{Timosh} is great.

The author is grateful to \  R.~I.~Bogdanov \ and \  V.~F.~Edneral \ 
for valuable discussions and \ E.~I.~Timoshkova \  for 
comprehensive commentary of~\cite{Timosh}. This work has been supported by 
the Russian Foundation for Basic Research under grants   
{\the\textfont2 
N}\lower-0.4ex\hbox{\small\underline{$\circ$}}~00-15-96560 and  
00-15-96577 and by  
the program "Universities of Russia".

\end{document}